\documentstyle[aps,prb,multicol,epsfig]{revtex}

\begin{document}

\title{Ferromagnetism and phase separation in one-dimensional \\ 
d--p and periodic Anderson models}
\author {M.\ Guerrero}
\address{
Theoretical Division, Los Alamos National Laboratory, Los Alamos, NM 87545}
\author{R.M.\ Noack}
\address{Institut de Physique Th\'eorique, Universit\'e de Fribourg,
CH--1700 Fribourg, Switzerland}
\date{ April 14, 2000}

\maketitle
\begin{abstract}
Using the Density Matrix Renormalization Group, we study metallic
ferromagnetism in a one--dimensional copper--oxide model which contains
one oxygen $p$--orbital and one copper $d$--orbital.
The parameters for the $d$--$p$ model can be chosen so that it
is similar to the one--dimensional periodic Anderson model.
For these parameters, we compare the ground--state phase
diagram with that of the Anderson model and find a
ferromagnetic region analogous to one found in the Anderson model, but
which is pushed to somewhat higher densities and interaction strengths.
In both models, we find a region within the ferromagnetic phase in which phase
separation between a localized ferromagnetic domain and a weakly
antiferromagnetic regime occurs.
We then choose a set of parameter values appropriate for copper--oxide
materials and explore the ground--state phase diagram as a function of the
oxygen--oxygen hopping strength and the electron density.
We find three disconnected regions of metallic
ferromagnetism and give physical pictures of the three different
mechanisms for ferromagnetism in these phases.

\end{abstract}


\begin{multicols}{2}
\narrowtext

\section{Introduction}

The nature of the microscopic description of metallic ferromagnetism,
as found in transition metals such as iron, nickel and cobalt, is a
long-standing problem in strongly correlated electron systems.
The Hubbard model was formulated in the 1960's \cite{hubbardetc} in
order to describe ferromagnetism in such materials.
However, a ferromagnetic ground state 
in the single--band Hubbard model on nonfrustrated bipartite lattices
has not been found at physical parameter values, and can be excluded
in a large portion of the ground--state phase diagram.
While mean--field theory yields a large region of stable
ferromagnetism, fluctuations tend to destabilize the ferromagnetic
phase, and numerical and variational calculations have narrowed the
possible extent of a ferromagnetic state to a small region around the
Nagoaka point.
Therefore, it has become clear that additional features must be added
to provide a description of metallic ferromagnetism.\cite{fazekas,vollhardt}

Recently, a number of such possible extensions to the 
Hubbard model have been investigated.
In particular, there are three classes of additions which can
enhance ferromagnetism: {\it (i)} a change in the noninteracting
density of states through the addition of frustrating
hopping terms or the treatment of a geometrically frustrated lattice
{\it (ii)} the inclusion of multiple orbitals per site or unit cell,
and {\it (iii)} the addition of more general nearest--neighbor interactions. 
\cite{fazekas,vollhardt}
In relation to {\it (i)}, Vollhardt et al. \cite{vollhardt} have
emphasized that peaks or singularities in the occupied portion of the
noninteracting density of states minimizes the kinetic energy loss due
to polarization of the electrons.
This picture is supported by numerical calculations in one
\cite{daul} and two dimensions,\cite{hlubina} and  within the Dynamical
Mean-Field Theory with a number of different forms of the
noninteracting density of states. \cite{vollhardt}
In addition,
Mielke \cite{mielke} has shown rigorously that a Hubbard model with a
less than half--filled flat band has a fully polarized ground state.
This theorem has been extended to the case of nearly flat
bands. \cite{mielketasaki}
For {\it (ii)}, a number of multiband models, such as the periodic
Anderson model, the Kondo lattice model, and the multiband Hubbard
models with Hund's rule coupling have been found to have
ferromagnetic ground states.\cite{fazekas} 
In relation to {\it (iii)}, Strack and Vollhardt \cite{strack} have
studied a quite general form of the nearest--neighbor interaction and
have found a fully polarized ground state at half--filling and one
hole from half--filling (the Nagaoka state) under certain conditions.

In this paper, we consider a one--dimensional model which
contains elements of points {\it (i)} and {\it (ii)} from above.
The model which we will study is a Cu--O chain that
contains one oxygen $p$--orbital and and copper $d$--orbital.
Since one--dimensional models can
usually be treated using well-controlled 
analytical or numerical techniques, one can obtain a complete
picture of the ground--state phase diagram without resorting to
approximations necessary in two or three dimensions.
For one--dimensional systems with nearest--neighbor hopping and an
arbitrary, but real and particle--symmetric interaction, a
theorem by Lieb and Mattis \cite{lieb} rules out a magnetized
ground state.
Because the Lieb--Mattis theorem no longer applies when electrons can
pass around each other, 
the minimal addition to the one--dimensional Hubbard model necessary
to obtain ferromagnetism is a next--nearest--neighbor hopping.
The resulting Hubbard model on a ``zigzag'' ladder has a ferromagnetic
ground state in a substantial parameter regime. \cite{daul}
One can also formulate more general, related models composed of
triangular elements.
Two such models, studied by Tasaki \cite{tasaki} and Penc et
al. \cite{penc} have been found to have ferromagnetic phases.
The $d$--$p$ model we study here is (aside from an added
near--neighbor Coulomb repulsion) a particular case of the model
studied in Ref.\ \onlinecite{penc}.

Another approach to ferromagnetism in one--dimensional models,
involving point {\it (ii)} from above, are models with multiple
non-degenerate orbitals per site.
The simplest of these models are the one--dimensional Periodic
Anderson Model (PAM) and the Kondo Lattice model (KLM).
Both of these models have ferromagnetic ground states in a relatively
large parameter regime. \cite{us,tsunetsugu}
These models can be related to one another in the regime of large
$f$--repulsion for the PAM and small Kondo coupling for the KLM.
The PAM is similar to the $d$--$p$ model in that there are two linked
non-degenerate levels per unit cell, and, as we will discuss in more
detail in the following, exhibit similar behavior with the
appropriate choice of parameters.

Another motivation for studying the $d$--$p$ model is the relationship
with models for the CuO$_2$ planes of the high--$T_c$ superconductors.
One dimensional $d$--$p$ models have been thought to contain some of
the essential ingredients needed to describe superconductivity in
CuO$_2$ planes. \cite{varma}
Because of this, previous work on the one--dimensional $d$--$p$ model
has concentrated on determining the conditions under which 
superconducting correlations are dominant for models with \cite{sano1}
and without \cite{sudbo,sano2,grilli} direct hopping between the $p$--orbitals.

In this work, we will concentrate on the ferromagnetic phases in the
ground state phase diagram and will discuss their physical origin.
In Sec.\ \ref{MODEL}, we introduce the models which we will study and
discuss choices of the models parameters.
In Sec.\ \ref{ANDERSON}, 
we compare the phase diagram of the Anderson 
lattice model from Ref.\ \onlinecite{us} with that of the $d$--$p$
model. 
We find that the ferromagnetic region persists in the $d$--$p$ model
but it is pushed to  higher values of the interaction and higher
fillings.  
There is a region of phase 
separation in which ferromagnetic and antiferromagnetic states coexist.
In Sec.\ \ref{REALISTIC}, we set the parameters of the $d$--$p$ system
to realistic values for the cuprates. 
We then investigate the phase diagram as a function of the
oxygen--oxygen hopping amplitude, $t_{pp}$, and
the band filling, and find three disconnected ferromagnetic regions
with three physically different mechanisms.

\section{\label{MODEL} Models}
\begin{figure}
\begin{center}
\epsfig{width=7cm,file=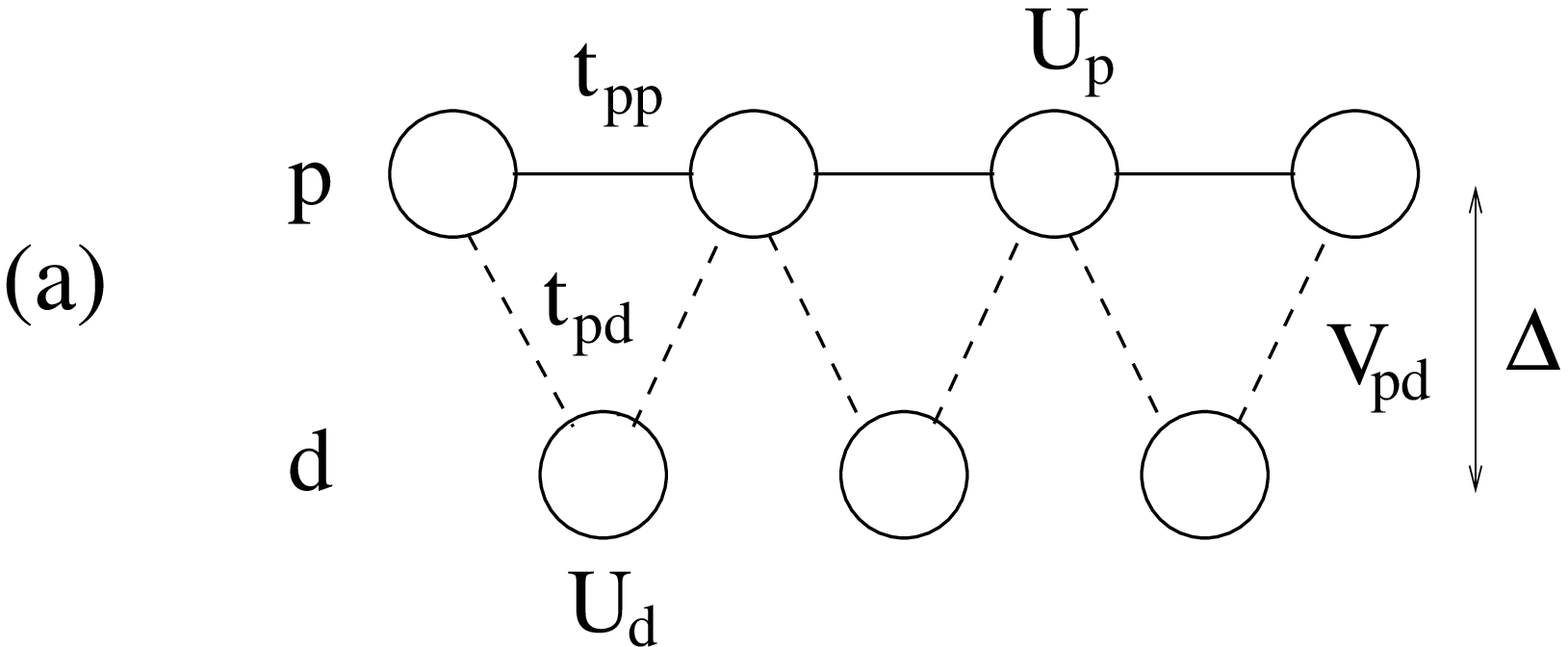}

\vspace*{0.3cm}
\epsfig{width=5.6cm,file=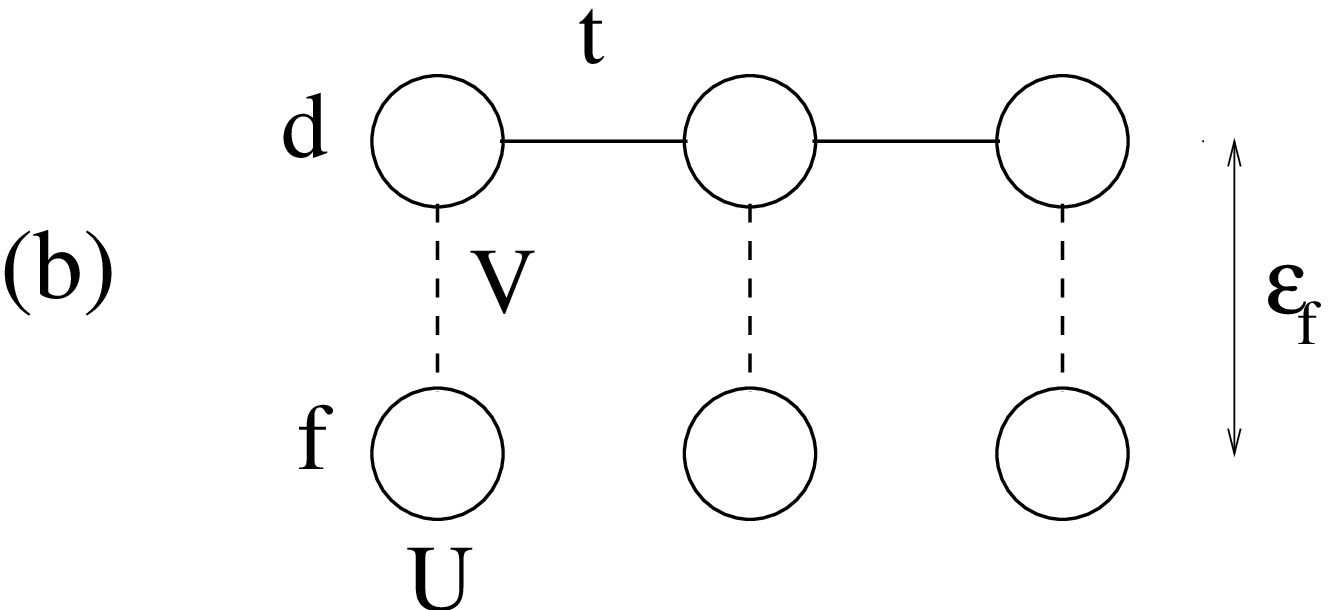}\hspace*{1.1cm}
\end{center}
\caption{A schematic diagram of (a) the copper--oxide lattice and (b)
the periodic Anderson model, with the system parameters marked. 
}
\label{fig01} 
\end{figure}

We consider a one-dimension $d$--$p$ system with Hamiltonian
\end{multicols}
\widetext
\begin{eqnarray}
 H = & -t_{pp} & \sum_{[jj^\prime] \sigma} (p^{\dagger}_{j \sigma}   
p^{\,}_{j^\prime\sigma} + p^{\dagger}_{j^\prime \sigma} p^{\,}_{j \sigma} ) 
 + \Delta \sum_{j \sigma} n^{p}_{j \sigma} 
 + U_p \sum_{j} n^{p}_{j \uparrow}n^{p}_{j \downarrow} \nonumber  \\       
& + & U_d \sum_{i} n^{d}_{i \uparrow}n^{d}_{i \downarrow}   
-t_{pd} \sum_{<ij>\sigma} (d^{\dagger}_{i \sigma} p^{\,}_{j \sigma}
 +  p^{\dagger}_{i \sigma}  d^{\,}_{j \sigma} )    
+ V_{pd}\sum_{<ij>}n^d_{i}n^p_j \; ,
\label{eqnCuOHam}
\end{eqnarray}
\begin{multicols}{2}
\narrowtext
\noindent
where $[jj^\prime]$ denotes a sum over nearest--neighbor oxygen pairs, and
$\langle i j \rangle$ a sum over copper--oxygen nearest neighbors. 
In the $d$--$p$ system we work in the hole representation so that
$d^\dagger_{i,\sigma}$ $(p^\dagger_{j,\sigma})$ creates a hole on
$d$ ($p$) site $i$ ($j$) with spin $\sigma$, and 
$n^{d}_{i \sigma}= d^{\dagger}_{i\sigma}d^{\;}_{i\sigma}$  
and $n^{p}_{j \sigma}$ are the local hole densities on a copper and oxygen
site, respectively.
The parameter $t_{pp}$ is the strength of the direct hopping between
the $p$-orbitals, $\Delta$ is the 
difference in on--site energies, $U_p$ and $U_d$ are the on--site Coulomb 
repulsion on the $p$ and $d$ sites respectively, $t_{pd}$ is the hybridization 
and $V_{pd}$ the Coulomb repulsion between nearest--neighbor $p$ and 
$d$ orbitals.
The lattice structure of the $d$--$p$ model including a schematic
representation of the Hamiltonian parameters is shown in 
Fig.\ \ref{fig01}(a). 
Unless otherwise stated, we will consider lattices consisting of $N$
copper sites and $N+1$ oxygen sites, with open boundary conditions at
the ends of the lattice so that the ends consist of oxygen sites with
connections in only one direction.
When the system has $N_{h}$ total holes, we will discuss 
$N_p \equiv N_{h} - N$, which corresponds to the excess number of
holes in the oxygen band when $U_d > \Delta > 0$,  and the corresponding
$p$--band filling, $n_p \equiv N_p/N$.

This model has a gauge symmetry which can be used to permute the signs
of the hopping matrix elements around each triangular element.
For example, alternating the sign of $t_{pd}$ on succesive bonds
and taking $t_{pp}$ positive is equivalent to having the same sign of
$t_{pd}$ and taking $t_{pp}$ negative. 
Here we take the latter case, as shown in Eq.\ (\ref{eqnCuOHam}),
restrict $t_{pd}$ to be positive, and allow the sign of $t_{pp}$
to vary.

In order to understand the physics of ferromagnetism in the $d$--$p$
model, it is useful to make a comparison with a model that has a
similar structure in an appropriately chosen parameter regime, 
the one--dimensional periodic Anderson model.
The PAM has Hamiltonian
\end{multicols}
\widetext
\begin{equation}
H=-t\sum_{i \sigma} (c^{\dagger}_{i \sigma}   c_{i+1 \sigma} +
                     c^{\dagger}_{i+1 \sigma} c_{i\sigma} )
  + \varepsilon_{f} \sum_{i \sigma} n^{f}_{i \sigma} 
  + U \sum_{i} n^{f}_{i \uparrow}n^{f}_{i \downarrow}
  + V \sum_{i \sigma} (c^{\dagger}_{i \sigma} f_{i \sigma} + 
                      f^{\dagger}_{i \sigma}  c_{i \sigma} ) \; ,
\label{eq:PAMHam}
\end{equation}
\begin{multicols}{2}
\narrowtext
\noindent
where $c^{\dagger}_{i \sigma}$ and $c_{i\sigma}$ create and annihilate
conduction electrons with spin $\sigma$ at lattice site $i$, and
$f^{\dagger}_{i \sigma}$ and $f_{i \sigma}$ create and annihilate
local $f$--electrons.
Here $t$ is the hopping matrix element for conduction electrons between
neighboring sites, 
$\varepsilon_{f}$ is the energy of the localized $f$--orbital,  
$U$ is the on--site Coulomb
repulsion of the $f$--electrons, and $V$ is the on--site hybridization
matrix element 
between electrons in the $f$--orbitals and the conduction  band.
We denote the number of electrons by $N_e$, and $N$ is the
number of unit cells
(each consisting of one $f$ site and one conduction site) in
the lattice.
Since there are two electronic orbitals in each site, the quarter--filled case 
corresponds to $N_e=N$ and the half--filled case has 
$N_e=2N$.
In analogy to the $d$--$p$ model, we will discuss $N_c = N_e - N$, the
excess number of electrons in the conduction band when 
$U > - \varepsilon_f > 0$, and the
corresponding filling, $n_c \equiv N_c/N$.
Fig.\ \ref{fig01}(b) shows a schematic represention of the PAM.

One can set the parameters of $d$--$p$ model so that it is the same as
the PAM except that the $d$-orbital hybridizes with the two
nearest--neighbor copper atoms, while the PAM hybridizes only on--site.
In order to compare the two models, we
set $V_{pd}=U_p=0$, $t_{pp}=t$, $t_{pd}=V$, $\Delta = -\epsilon_f$ and  
$U_d = U$. 

\section{\label{RESULTS} Calculation and Results}

Since itinerant ferromagnetism is an intrinsically strong--coupling
phenomenon that occurs due to a subtle competition between kinetic and
potential energy, \cite{vollhardt} it is generally quite hard to treat
with analytic methods.
While exact statements can be made in some special cases, such as the
Nagaoka state \cite{nagaoka} and the case of flat bands,\cite{mielke}
in general one must resort to mean--field theory or variational
techniques, both of which give phase diagrams which can be at best
qualitatively accurate.
Therefore, sufficiently accurate numerical methods can be useful to
determine the properties of models with ferromagnetic phases.
Here we use the Density Matrix Renormalization Group to
calculate the ground--state properties of the $d$--$p$ model and the PAM.
The DMRG is a variational numerical method closely related to exact
diagonalization, but which can be used to treat much larger 
systems.\cite{white,DMRGbook}
Being able to treat sufficiently large systems is important since the
models we study here have two fermionic sites, i.e.\ sixteen degrees of
freedom per unit cell, which would severely limit the maximum size
available for exact diagonalization.
Quantum Monte Carlo methods would suffer from the fermion sign problem
at all band fillings on the $d$--$p$ lattice.

We use the finite--system version of the ground--state DMRG
algorithm \cite{white} to accurately calculate the energy expectation
values of equal--time operators in the ground state including local spin,
density, and various correlation functions.
We keep up to 800 states in the system block and treat lattices
of up to $N=32$ unit cells.
The maximum sum of discarded density matrix eigenvalues is
approximately $5\times10^{-6}$. 

In order to search for ferromagnetic phases, it is essential to be
able to determine the total spin of the ground state.
Since our version of the DMRG algorithm does not allow direct control
of the total spin, we use a combination of methods to do this.
First, we can calculate the expectation value 
$\langle \psi_0 | S^2 | \psi_0 \rangle$ directly, where $\psi_0$ is the
variational DMRG ground state calculated in a particular $S_z$ sector,
and $S^2$ is the explicit operator for the square of the total spin.
Since the DMRG ground state is variational,
$\langle \psi_0 | S^2 | \psi_0 \rangle$ does not always take on the
quantized values due to mixing between states of different $S_z$.
Second, we can add a term of the form $\lambda S^2$ to the
Hamiltonian, raising the energy of higher spin states within a
particular $S_z$ sector.\cite{daul}
We can use this to calculate the ground--state energy as a function
of $S^2$ by examining the energy of the minimum $S^2$ ground state in
a particular $S_z$ sector.
Finally, we can examine the ground--state energies as a function of
$S_z$ and use the degeneracy in $S_z$ to determine $S^2$.
For application of these methods to the ferromagnetism in the PAM, see
Ref.\ \onlinecite{us} and to the Hubbard chain 
with next--nearest--neighbor hopping, see Ref.\ \onlinecite{daul}.

\subsection{\label{ANDERSON} Comparison with the periodic Anderson model}

In the first part of this work, we present a comparison between the $d$--$p$ 
system and the PAM.
The qualitative picture developed for the
ferromagnetic phase of the PAM \cite{moeller,us} should also apply to the
$d$--$p$ model. 
In order to make the correspondence to the PAM,
we choose a regime in which the $d$-orbital lies below the
bottom of the $p$-band.
At exactly quarter filling, all the electrons are in the 
$d$-sites and all the spin states are degenerate if there is no
hybridization. 
When the hybridization
is turned on, the super-exchange favors antiferromagnetic interactions.
The exchange coupling 
constant can be calculated from perturbation theory giving
\begin{equation}
J_{A} = \frac{4t_{pd}^4}{(\Delta+V_{pd})^2} \left(
\frac{1}{U_d} + \frac{2}{2 \Delta + U_p}
\right) \; ,
\label{eqnJa}
\end{equation}
or when $U_d = 2\Delta$ and $U_p = V_{pd} = 0$, 
$J_{A} = 6t_{pd}^4/\Delta^3$. 
When a hole is added to the quarter--filled system, it will prefer
to go to the $p$-sites since $U_d>|\Delta|$. 
This extra hole will tend to form
singlets with the mostly localized spins in the d-sites. 
The hole can delocalize and thus lower its energy if 
the spins of the $d$--holes are oriented 
in the same direction, i.e.\ are ferromagnetically ordered.
The binding energy for these singlets can be calculated within perturbation 
theory in a manner similar to that applied to the two-dimensional case
by Zhang and Rice, \cite{Zhang} yielding
\end{multicols}
\widetext
\begin{equation}
J_S = 2 t_{pd}^2\left( 
\frac{1}{\Delta - V_{pd}} + \frac{1}{\Delta - V_{pd} + U_p} 
+ \frac{2}{U_d - \Delta - V_{pd}} \right) \; . 
\end{equation}
\begin{multicols}{2}
\narrowtext
\noindent
For $U_d = 2 \Delta$ and $U_p = V_{pd} = 0$, 
$J_S = 8t_{pd}^2/\Delta$.

In general, when there
are $N_p = N_{h} - N$ holes in the $p$-band, this effect will favor a
ferromagnetic ground state with total spin $S=(N-N_p)/2$. 
We will denote this value of $S$
for the ground state as {\it complete} ferromagnetism, while
a value of $S$ smaller but still greater than the minimum, will be
labeled {\it incomplete}.
Note that a complete state has a lower magnetization than a saturated one.
It represents a state where all the {\it uncompensated} d-spins are
aligned.

Thus, one would expect a competition between ferromagnetism and
antiferromagnetism near quarter filling similar to that found in the
PAM.\cite{us}
For the PAM there is an exact result that proves the ground
state is ferromagnetic with complete magnetization
for $N_c = 1$ ($N_e = N+1$) and $U_d=\infty$.\cite{sigrist,yanagisawa}
There is no such result, however, for the $d$--$p$ model. 
Furthermore, while the 
antiferromagnetic exchange is a sixth--order process in the Anderson 
lattice model, it is fourth order (and therefore stronger)
in the $d$--$p$ system.  
Therefore, it is not clear
that the ferromagnetism will still be present in the $d$--$p$ system.
Yanagisawa \cite{yanagisawa} studied a $2 \times 2$ CuO cluster with
exact diagonalization and found a ferromagnetic ground state. 
However, it is difficult to draw general conclusions about the
behavior in the thermodynamic limit from such a small cluster.

\begin{figure}
\begin{center}
\epsfig{width=7cm,file=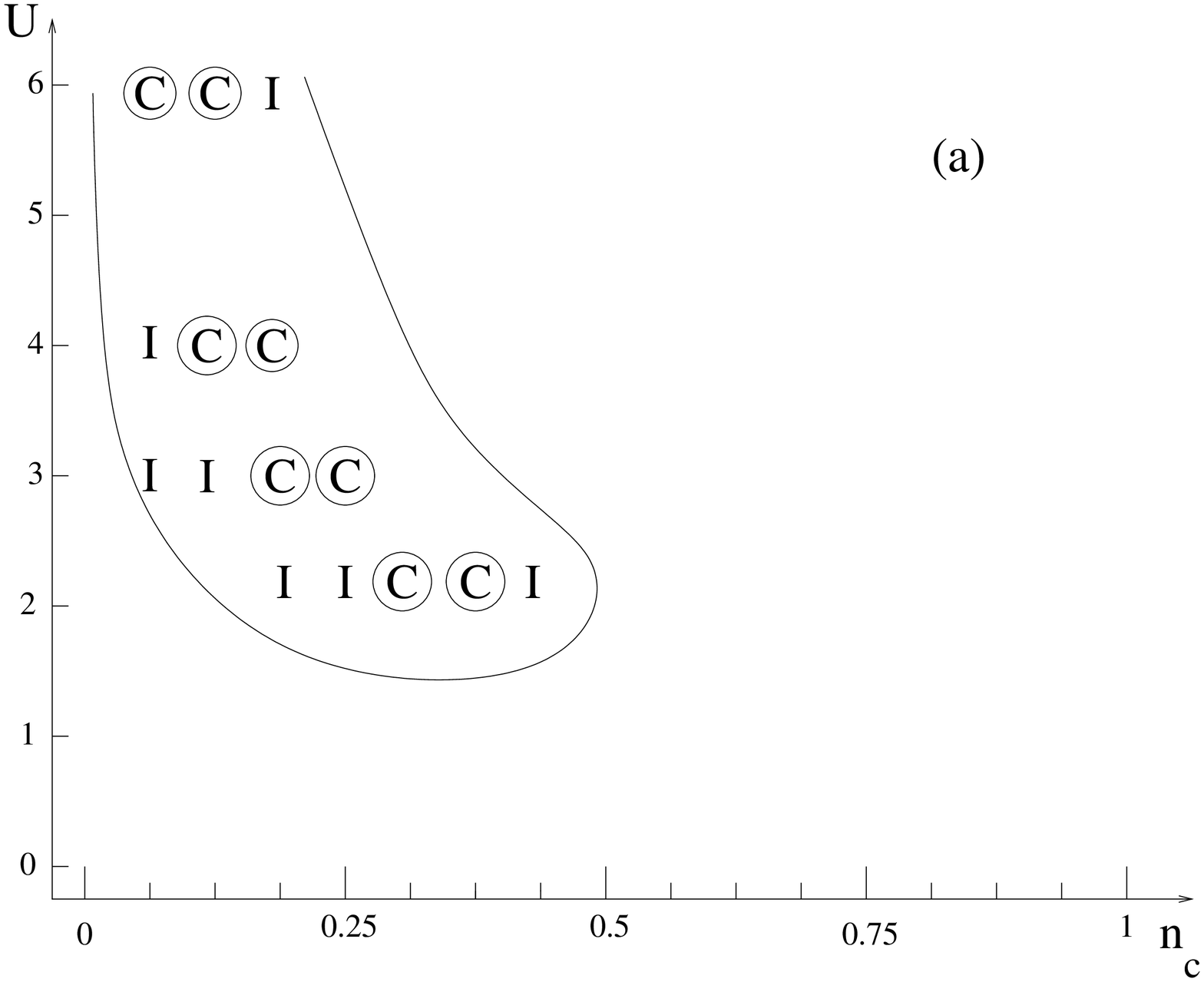}

\epsfig{width=7cm,file=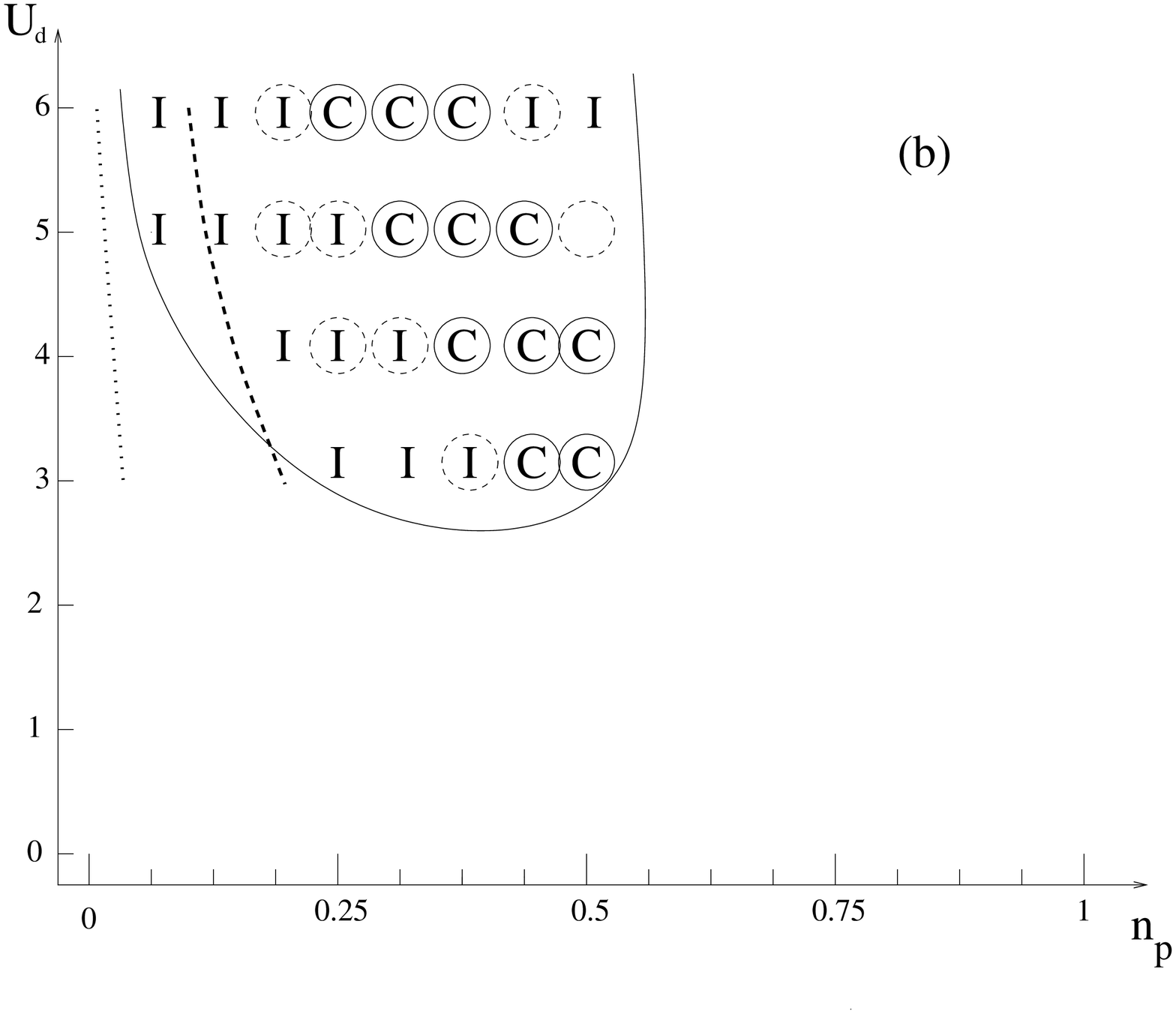}
\end{center}
\vspace*{-0.3cm}
\caption{
Ground--state phase diagram in the plane of band
filling and the local Coulomb repulsion for an $N=16$ system
of (a) the Anderson lattice with 
$V = 0.375$, $t=0.5$, $\epsilon_f= -U/2$, and 
(b) the copper--oxide lattice with equivalent parameters:
$t_{pd}=0.375$, $t_{pp}=0.5$, $\Delta=U_d/2$, $U_p = V_{pd} = 0$.
At $n_c=0$ or $n_p=0$, each system is quarter--filled, i.e.\ has one
hole per unit cell.
The dotted and dashed lines in (b) correspond to
Eq.\ (\protect\ref{eqnnpc1}) and Eq.\ (\protect\ref{eqnnpc2}),
respectively.
}
\label{fig02} 
\end{figure}

In our DMRG calculations, we consider chains with $N=16$
and open boundary conditions and set $t_{pp}=0.5$, $t_{pd}=0.375$,
$\Delta= U_d/2$ and $U_p = V_{pd} = 0$. 
This choice allows us to compare the phase diagram 
with the one of the Anderson lattice model from Ref. \onlinecite{us}
where $t=0.5$, $V=0.375$, $\epsilon_f = -U/2$. 
In Fig.\ \ref{fig02}, we
present both phase diagrams in the plane of band filling and the local 
Coulomb repulsion, $U$ or $U_d$. 
Here $n_c=0$  ($n_p =0$ for the $d$--$p$ model) represents the
quarter-filled case, which in
the large Coulomb repulsion regime means that the localized orbitals are 
singly occupied and the extended orbitals are empty in the zero
hybridization limit.
In the $d$--$p$ system, Fig.\ \ref{fig02}(b), the region of complete
ferromagnetism is pushed towards higher values of $U_d$ and higher
densities and there is a wide region of incomplete states at low
densities. 
This is consistent with the 
fact that antiferromagnetic correlations at quarter filling in this
system are of lower order (and therefore stronger) than in the
Anderson lattice model.

\begin{figure}
\begin{center}
\epsfig{width=6.5cm,angle=270,file=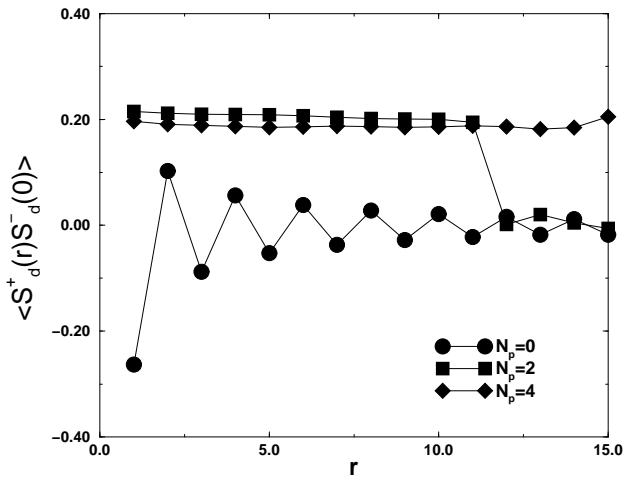}
\end{center}
\vspace*{-0.3cm}
\caption{ The Cu--Cu spin--spin correlation function as a function of
distance, $r$, for the copper--oxide lattice with the same parameters as in 
Fig.\ \ref{fig02} and $U_d = 5$ for 3 different fillings. 
}
\label{fig03} 
\end{figure}

In Fig.\ \ref{fig03} we show the Cu--Cu correlation function, 
$\langle S^+_{d}(r)S^-_{d}(0) \rangle$,
as a function of distance, $r$ (measured in units of the
lattice constant), for the $d$--$p$ model with
the parameters described above and $U_d=5$ for 
$N_p=0$, 2 and 4. 
Here $S^+_{d}(r) = d^\dagger_{r,\uparrow} d_{r,\downarrow}$ is the
$d$--spin raising operator on site $r$.
For $N_p = 0$ (quarter filling),
the correlations are clearly antiferromagnetic with the amplitude decaying
slowly with distance, while for $N_p=2$ and $N_p=4$, the correlations
are ferromagnetic.
This illustrates that the system goes from an antiferromagnetic state
at quarter filling to a ferromagnetic state as soon as additional
holes are added for these parameter values.

The competition between the tendency to ferromagnetism and the
antiferromagnetic  
exchange gives rise to the region of {\it incomplete} ferromagnetism
at low doping.
In this region, we find that the system {\it phase separates} into domains 
of complete  ferromagnetism in which the $p$--holes are localized and 
antiferromagnetic domains in which the $p$--band is empty. 
Indeed, the first symptom of this phase
separated state can be seen in Fig.\ \ref{fig03}, in which there is a
jump in the correlation function for $r=12$ for the case of 18 
holes. 
This peculiar 
behavior is due to the tendency of the holes to localize and form a
region of complete ferromagnetism near the center of the chain, leaving
the p-orbitals empty near the ends.
	
\begin{figure}
\begin{center}
\epsfig{width=6.5cm,angle=270,file=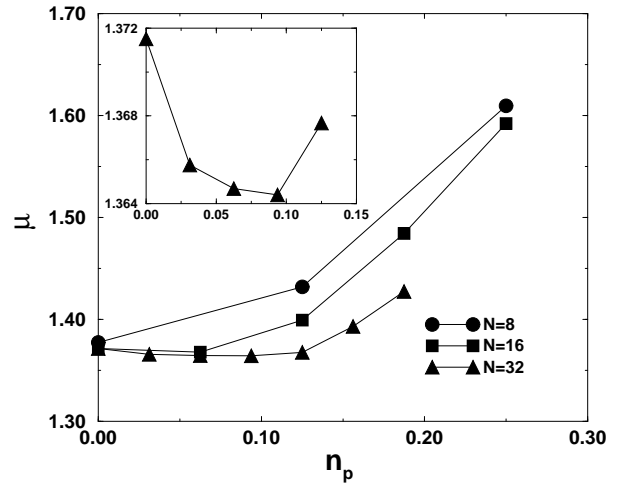}
\end{center}
\caption{The chemical potential $\mu$, as a function of the $p$--band
filling, $n_p$,
for the Cu-O lattice with the same parameters as in Fig.\ \ref{fig03}
for different system sizes.
The inset is an expanded view of the low--density results for the 32
unit--cell lattice showing a minimum in the chemical potential, the
signature for phase separation.
}
\label{fig04} 
\end{figure}

To illustrate this phase--separated
regime, we have calculated the chemical potential $\mu = E_o(N_h+1)-E_o(N_h)$, 
where $E_o(N_h)$ is the ground state energy with $N_h$ holes,
for chains of $N=8$, 16 and 32 unit cells. 
We show $\mu$ as a function of the filling for $U_d=5$ in Fig. \ref{fig04}. 
For 8 unit cells, $\mu$ increases monotonically with $n_p$, but for 16
unit cells there is a shallow minimum at $n_p = 0.0625$. 
For 32 unit cells, there is a flat region which is shown
in detail in the inset, with a minimum at $n_p=0.09375$
which is indicative of phase separation.
We find a similar behavior of the chemical potential is found for
$U_d=6$ and $U_d=4$. 

\begin{figure}
\epsfig{width=6.8cm,angle=270,file=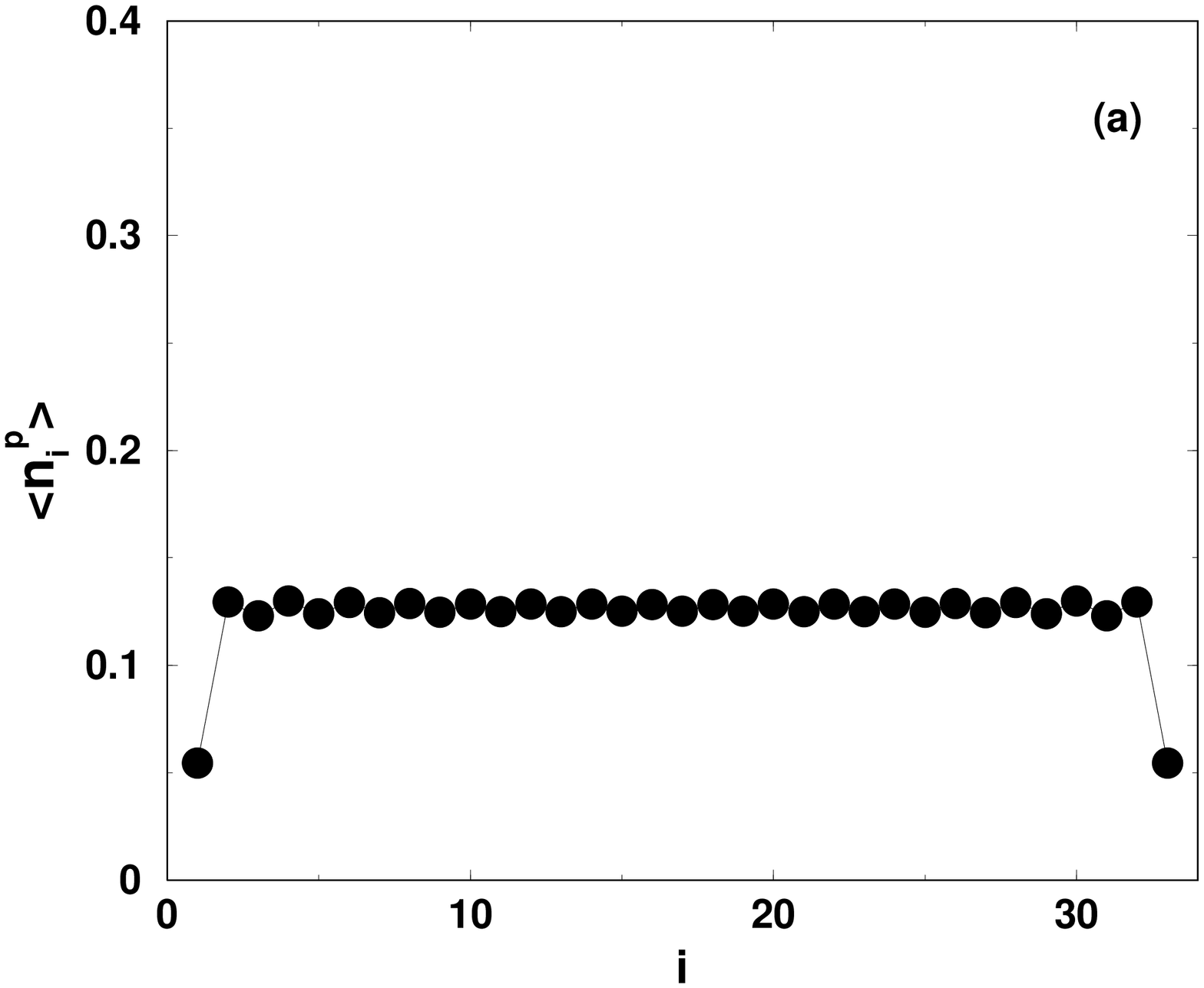}
\hfill
\epsfig{width=6.8cm,angle=270,file=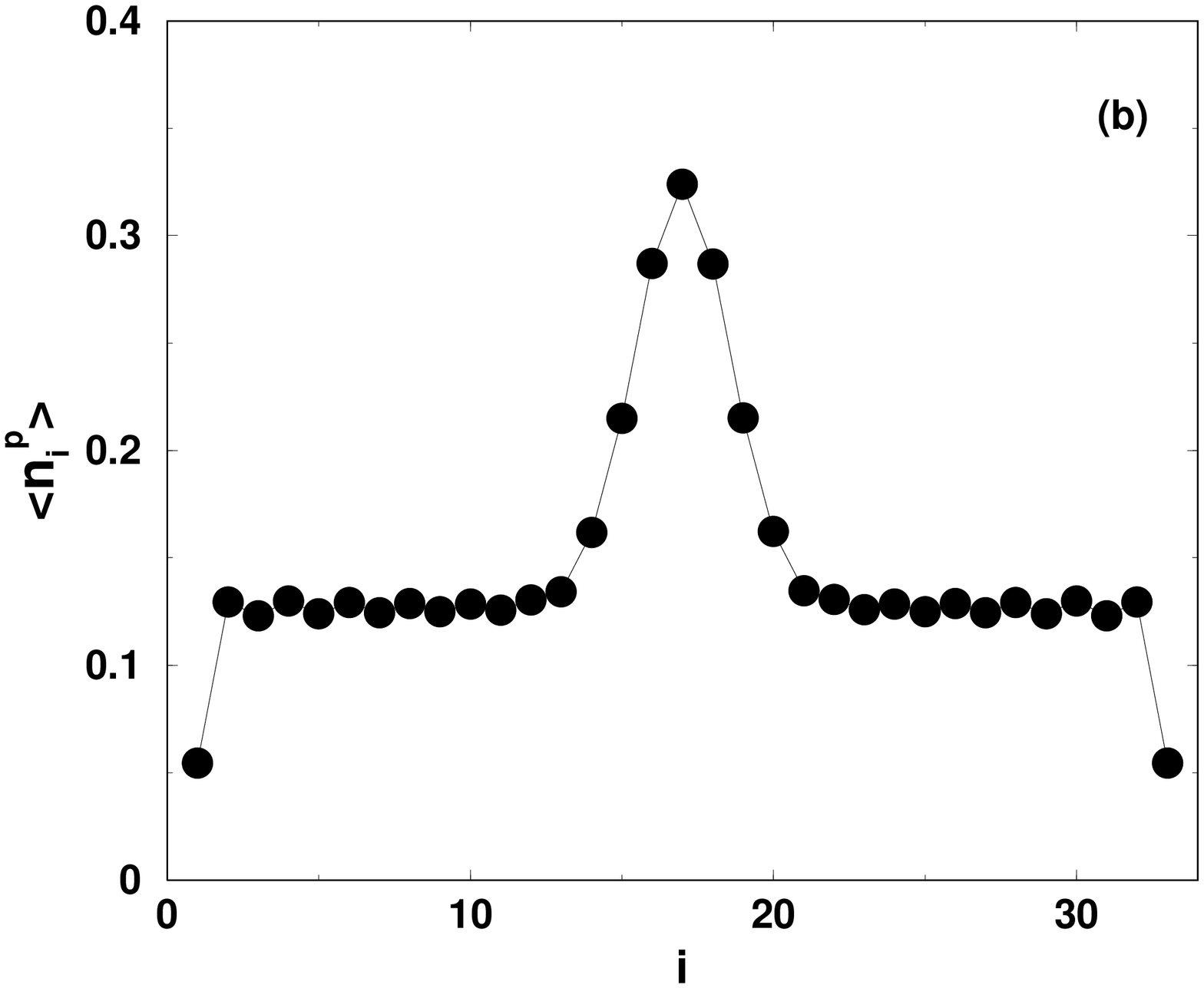}\hspace*{0.6cm}

\epsfig{width=6.8cm,angle=270,file=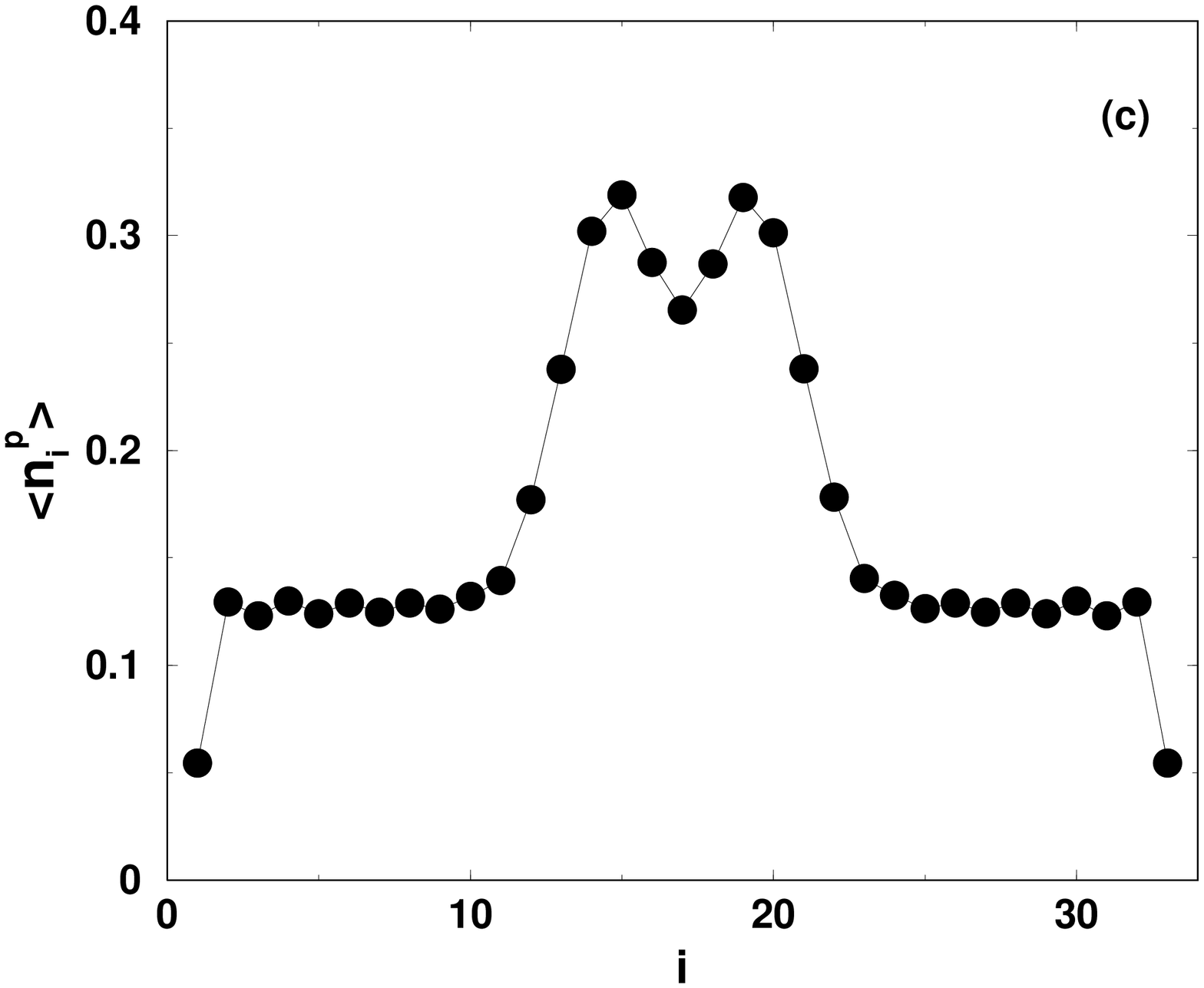}
\hfill
\epsfig{width=6.8cm,angle=270,file=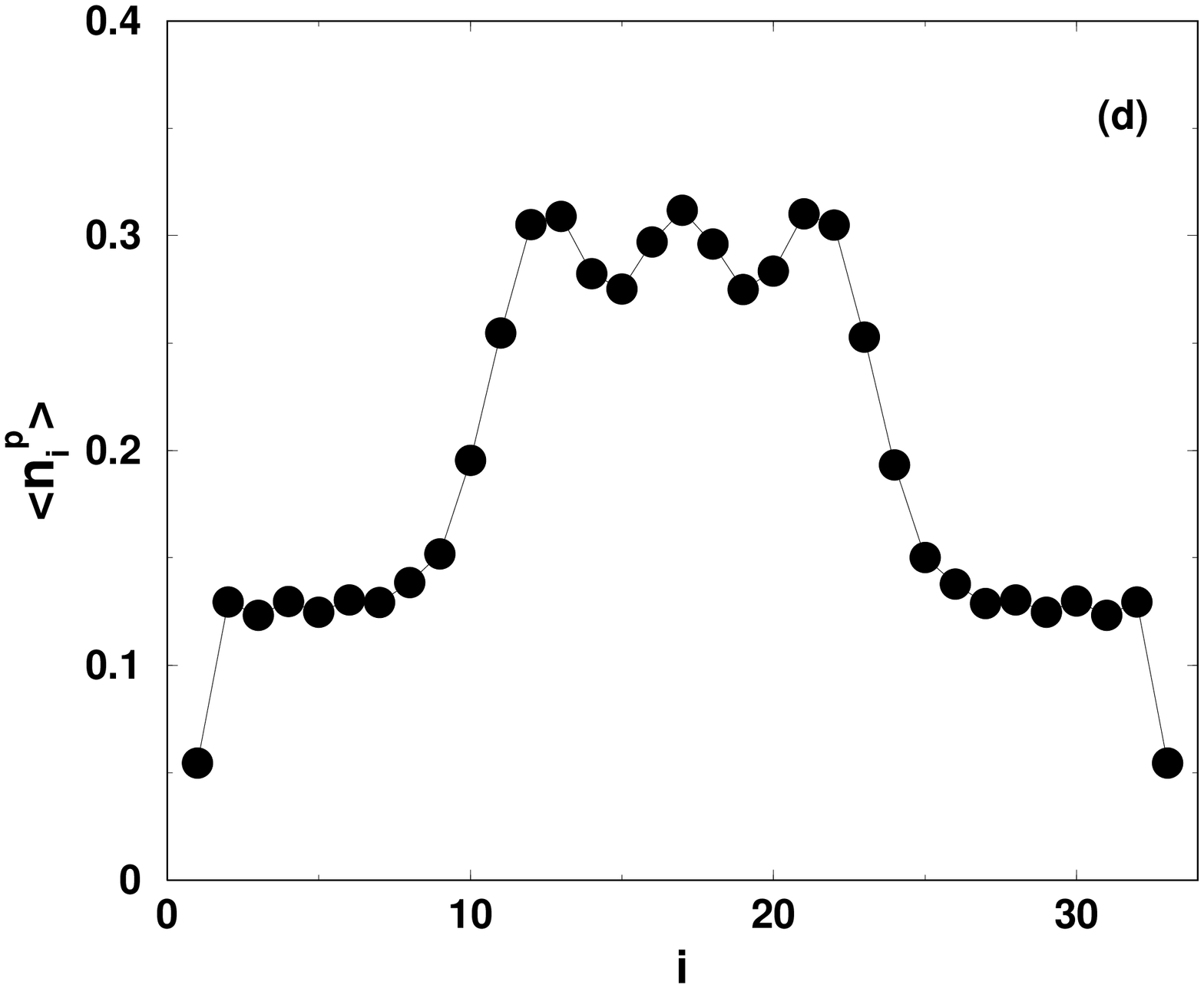}\hspace*{0.6cm}

\caption{Density in the oxygen sites as a function of position in the lattice,
$i$, with $N=32$ unit cells, the same parameters as in Fig. \ref{fig02}
and $U_d = 4$ for (a) $N_p=0$ (quarter filling), (b) $N_p=1$, 
(c) $N_p=2$, and (d) $N_p=3$.
}
\label{fig05} 
\end{figure}

To illustrate the localization of the holes in the $p$--band, we
display the density in the $p$--sites as a function of position in the
lattice for $U_d=4$ in an $N= 32$ unit--cell chain with $N_p = 0,1,2,3$
in Fig.\ \ref{fig05}.
At quarter filling, the $p$--levels are almost empty. 
Upon doping, they begin to fill with holes, but
instead of spreading out over the lattice, the additional holes
localize in a region which is smaller than the lattice size.
This behavior is also seen for $U_d=5$ and $U_d=6$.

\begin{figure}
\epsfig{width=6.5cm,angle=270,file=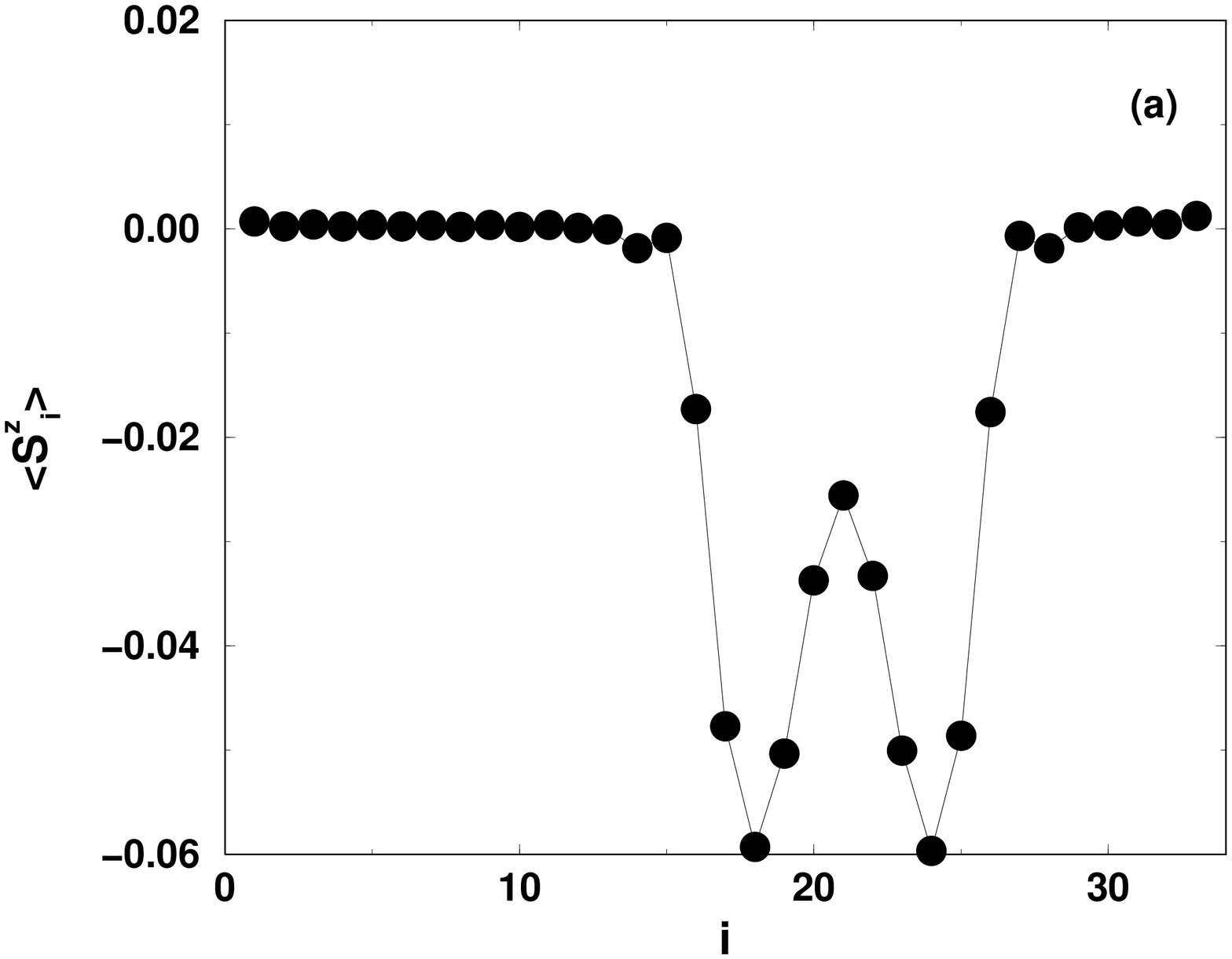}
\hfill
\epsfig{width=6.5cm,angle=270,file=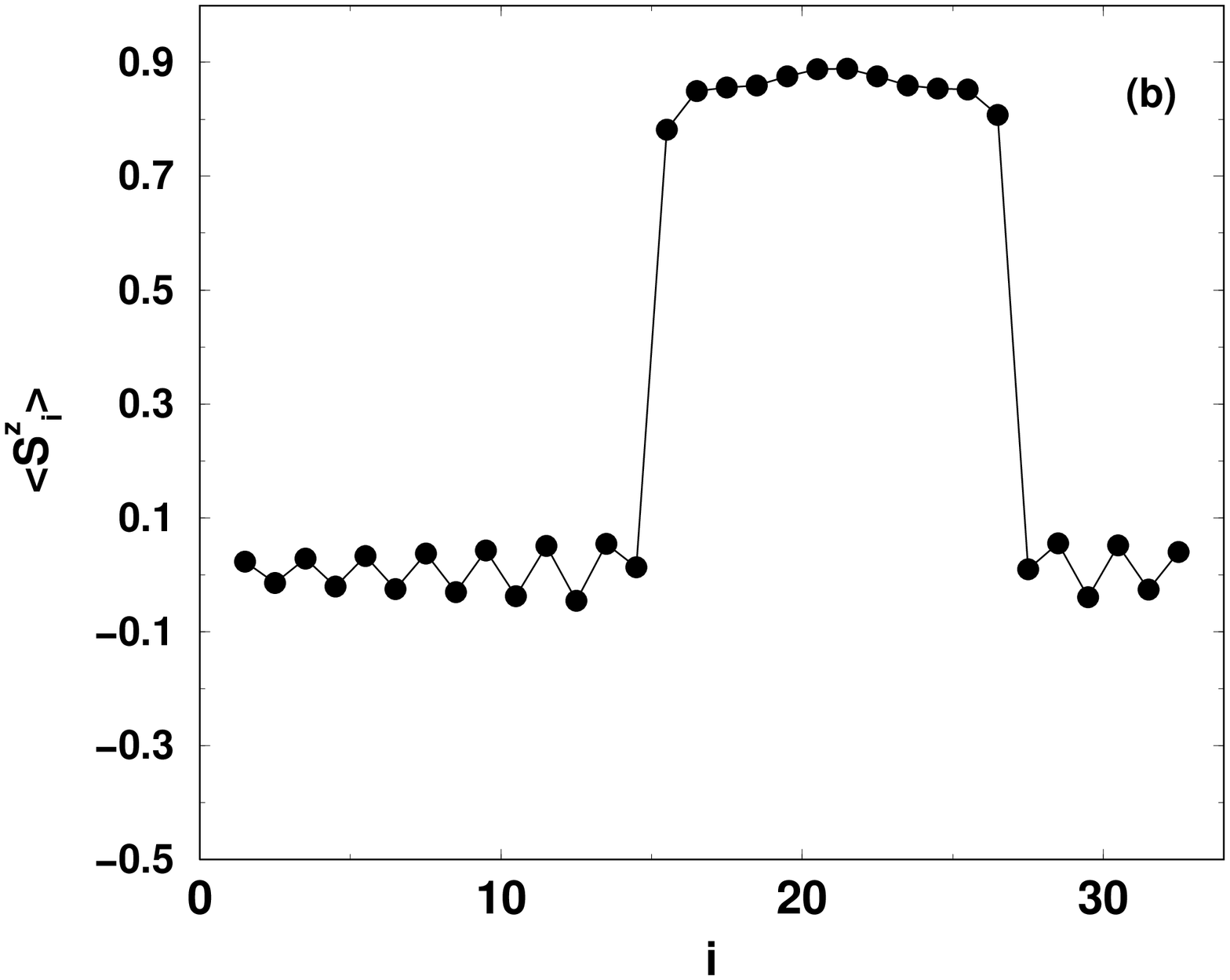}\hspace*{0.5cm}
\caption{The local spin density $\langle S_i^z \rangle$ as a function of
lattice position, $i$, for (a) the oxygen sites and (b) the copper
sites on an $N= 32$ lattice with $S_z = S = 5$ (the ground state) and
$N_p=2$.
Here $U_d = 5$ and the remaining parameters are the same as in
Fig.\ \ref{fig02}.  
}
\label{fig06} 
\end{figure}

In Fig.\ \ref{fig06} we show the local spin density 
$\langle S_i^z \rangle$ as a function of position in 
the lattice for an $N=32$ chain with $U_d=5$, $N_p=2$ and
$S_z=S=5$ (the ground state) for both the $p$--band and the
$d$--band. 
There are two distinct domains in the lattice:
one in which the $d$--holes are completely polarized and another
one in which they tend to order antiferromagnetically. 
The coupling between $p$ and $d$ holes is clearly antiferromagnetic.
The asymmetry of the distribution is a numerical artifact due to the
near--degeneracy of states with the phase--separated region centered
at different points.

To gain some physical insight into the phase separation mechanism,
one can use energetic arguments to estimate the stability of competing
phases.
The energy scale in the antiferromagnetic state is set by $J_A$, the
antiferromagnetic exchange energy, Eq.\ (\ref{eqnJa}).
The energy of the uniform ferromagnetic phase can be estimated by the
binding energy of the Zhang--Rice--like singlets, $J_S N_p$.
For a rough estimate of the magnetic energy scale, consider 
an antiferromagnetically ordered state versus one that has complete
ferromagnetism.
The Bethe Ansatz energy of a one--dimensional Heisenberg
antiferromagnet \cite{bethe} consisting of $N-1$ bonds (measured
relative to the ferromagnetic state and neglecting end effects) is 
$J^{\rm eff}_A (N-1)$, where  
$J^{\rm eff}_A = \ln 2 J_A$.
In order for complete ferromagnetism to occur,
the condensation energy of the singlets, 
$J_S N_p$ must be larger than this, leading to
the condition 
\begin{equation}
n_p \gtrsim \ln 2 J_A/J_S \equiv n_{pc1} \; .
\label{eqnnpc1}
\end{equation}
This estimate is only valid for very low densities of $p$--electrons,
since it neglects their kinetic energy, and for strong coupling where
the magnetic picture is valid.
According to this estimate, ferromagnetism is favored over
antiferromagnetism for $U_d \gtrsim 3$ at all densities above
quarter-filling on the lattice sizes that we have studied.
However, it yields a smaller $n_{pc1}$ than we find in the numerical
calculations.
For example, for $U_d=4$ and the 
parameter set used in this section, $n_{pc1} = 0.018$.

\end{multicols}
\widetext

\begin{multicols}{2}
\narrowtext

Now let us examine the possibility of phase separation between a
region of undoped antiferromagnetism and a region of complete
ferromagnetism.
In order for such a phase to exist, the energy gained by forming an
undoped antiferromagnetic domain must compensate the loss in kinetic
energy of the Zhang--Rice singlets, which have to localize in a region
smaller than the lattice size $N$.
We estimate the kinetic energy of the Zhang--Rice singlets by taking
them to be noninteracting hard--core bosons confined to a lattice of
length $X$ with open boundary conditions.
(We study lattices with open boundary conditions here,
so that this form is valid up to the number of $p$--sites, $X=N+1$.)
The single--particle energy levels are given by
\begin{equation}
\varepsilon_j = -2 t_{ZR} \cos \frac{\pi j}{X+1} \; ,
\end{equation}
where $t_{ZR}$ is the effective hopping of the singlets and 
$j=1,\ldots,N$.
Since hard--core bosons are equivalent to spinless fermions in one
dimension, the total kinetic energy is given by successively
occupying single--particle states,
\begin{eqnarray}
T(N_p, X) & = & \sum_{j=1}^{N_p} \varepsilon_j \nonumber \\
& \approx & \int_0^{N_p} dj \;
\varepsilon_j
\approx -2 t_{ZR} \; \frac{X}{\pi} \; \sin \frac{N_p \pi}{X} \; ,
\label{eqnKE}
\end{eqnarray}
where the approximation on the right--hand side is valid in the
continuum limit: $X \gg 1$.
The energy gained by constricting the ferromagnetic domain to a length
$X$ is
\begin{eqnarray}
E_{PS} (N_p,X) = & \ & J^{\rm eff}_A (N-1-X) \nonumber \\ 
& + & T(N_p,N+1) - T(N_p,X) \; .
\label{PS1}
\end{eqnarray}
The phase--separated state will be favored over the homogeneous
ferromagnetic state if $E_{PS}$ is positive.
In that case, the size of the ferromagnetic region is given by the
value of $X$ that maximizes the energy gain, determined by
\begin{equation}
0 = 
\frac{\partial E_{PS}}{\partial X} \approx
- J^{\rm eff}_A + \frac{2 \pi^2 t_{ZR} N_p^3 }{3 X^3} \; ,
\end{equation}
where we have taken $N_p \pi \ll 1$ in order to expand the sine and
have used the continuum expression from the right--hand side of 
Eq.\ (\ref{eqnKE}).
Eq.\ (\ref{PS1}) has one maximum for $X >  0$ which occurs at 
\begin{equation}
X_{\rm max} 
= N_p \left( \frac{2 \pi^2 t_{ZR}}{3 J^{\rm eff}_A}\right)^{1/3}
\equiv N_p x_o \; .
\label{eqnPSlength}
\end{equation}
It is easy to verify that $E_{PS}(N_p,X_{\rm max})$ is always positive
as long as $0 <  X_{\rm max} < N$.
Therefore $X_{\rm max}$ is the length of the ferromagnetic domain when
phase separation is present.
Note that $X_{\rm max} \propto N_p$, in agreement with the behavior
seen in Figs.\ \ref{fig05} and \ref{fig06}.
In order to numerically estimate $x_0$ for the $d$--$p$
system, we take $t_{ZR} = t_{pp}$
and $J^{\rm eff}_A = \ln 2 J_A$, where $J_A$ is given
by Eq.\ (\ref{eqnJa}) for the $d$--$p$ model.
(One could calculate $t_{ZR}$ more accurately in perturbation theory,
but only its order of magnitude is important, since it appears within
the cube root.)
For the parameters used in this section, this gives 
$x_0 = 7$, 8.5 and 10 for $U_d=4$, 5 and 6, respectively. 
This is in reasonable agreement with our numerical results 
(See Fig. \ref{fig05}), which also show
an increase in the localization length with $U_d$.

\end{multicols}
\widetext

\begin{multicols}{2}
\narrowtext

The boundary between the uniform ferromagnetic phase and the
phase--separated phase is determined by the condition $X_{max} = N$,
leading to a critical density
\begin{equation}
n_{pc2} = \left( \frac{3 J^{\rm eff}_A}{2 \pi^2 t_{ZR}}\right)^{1/3}
\; .
\label{eqnnpc2}
\end{equation}
The densities $n_{pc1}$ and $n_{pc2}$ calculated
from Eqs.\ (\ref{eqnnpc1}) and (\ref{eqnnpc2}) are represented in the
phase diagram of  Fig.\ \ref{fig02}(b) as dotted and dashed lines,
respectively. 
Notice that the numerical calculation yields a
somewhat wider region of phase separation than given by these estimates.
In order to estimate the finite--size effects, we
have also examined the spin of the ground state on
$N=8$ and $N=32$  unit--cell chains.
We find that the points of complete polarization in 
Fig.\ \ref{fig02}(b) behave consistently with system size, i.e.\ are
also completely polarized for $N=8$ and $N=32$.
However, the scaling behavior of the incomplete states is more
complicated.
In particular, some states which are incompletely polarized in the $N=16$
chain become completely polarized for $N=32$.
These cases are circled with a dashed line in Fig. \ref{fig02}(b). 
Therefore, the phase--separated region tends to become narrower in the
thermodynamic limit.
There is, however, evidence that some of the incompletely polarized
states remain in the thermodynamic limit.
For $U_d=5$ and 6, the two densities closest to quarter filling on
the $N=16$ chain have total spin $S=5/2$ and
$5$, and $S=7/2$ and $6$, respectively.
On the $N=32$ chain, the ground--state $S$ at the same density 
doubles, suggesting that $S/N$ will remain constant in the
thermodynamic limit.
At $U_d=4$ and $U_d=3$, none of the incomplete states
show such a simple scaling and we believe that larger system sizes are needed
to determine the nature of the phase boundaries. 
For $U_d=3$, we find no evidence of phase separation at any finite
system size.

\begin{figure}
\begin{center}
\epsfig{width=6cm,angle=270,file=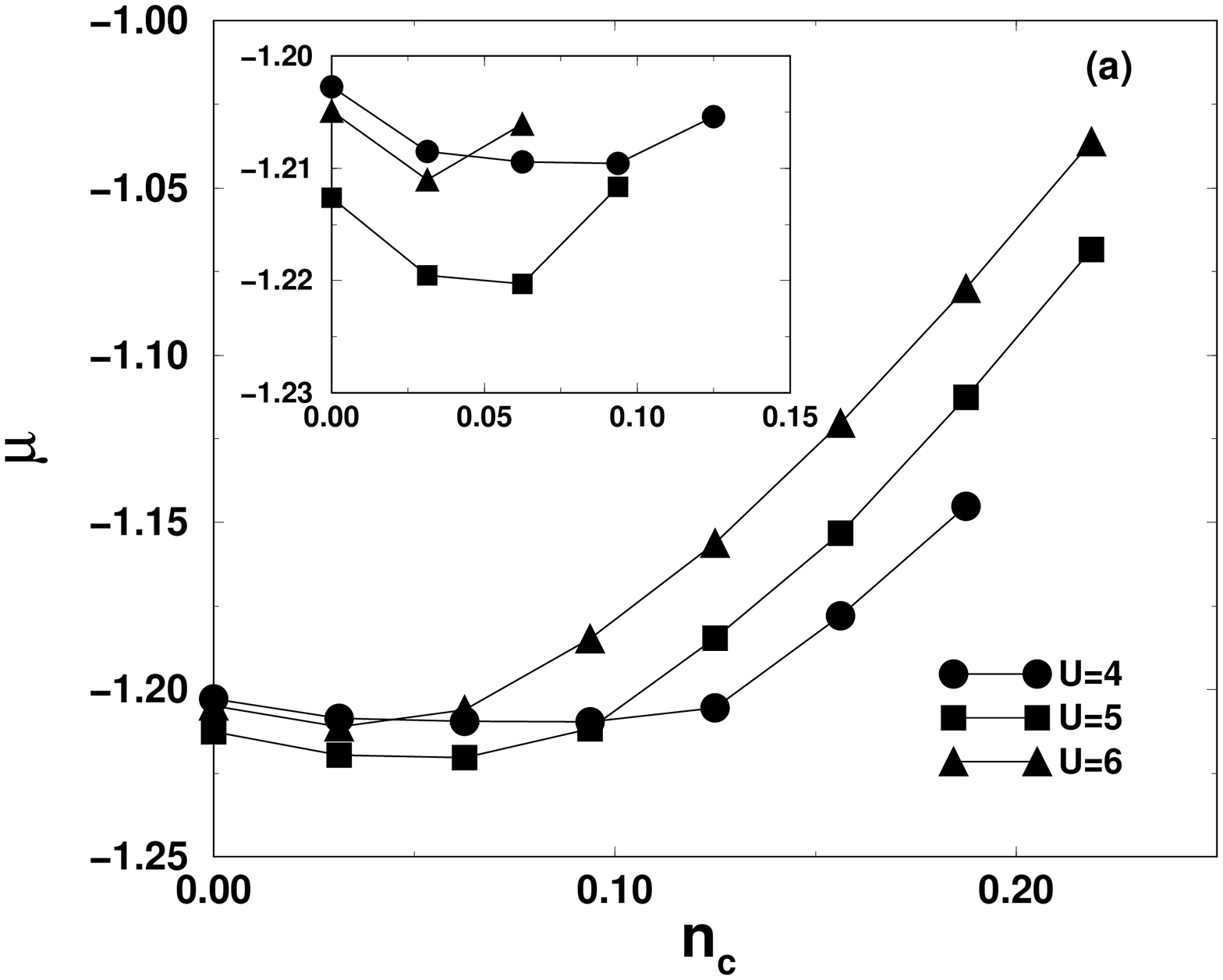}\hspace*{0.3cm}

\epsfig{width=6cm,angle=270,file=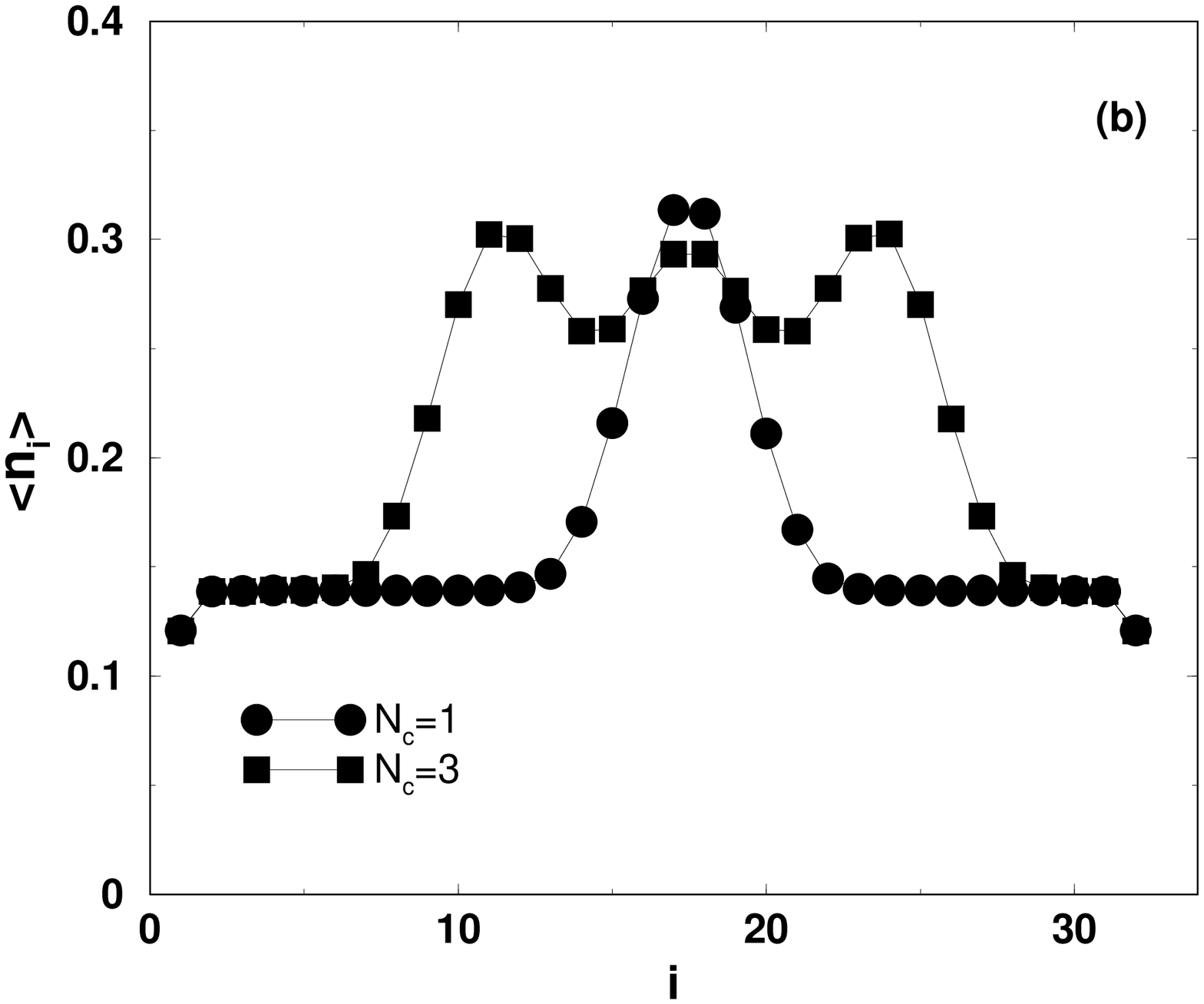}
\end{center}
\caption{ (a) The chemical potential, $\mu$, versus density, $n_c$, for
the periodic Anderson model with  $V = 0.75$, $t=0.5$, and 
$\epsilon_f= -U/2$ on an $N=32$ lattice. 
It shows a minimum for all three $U$--values.
(b) Conduction electron density versus site for the $U=5$ case, showing 
localization of the excess conduction electrons in a region smaller than the 
lattice size.
}
\label{fig07} 
\end{figure}

Let us now examine the relevance of the phase separation arguments to
the PAM.  
Our energetic arguments should also apply with the appropriate mapping
of $J_A$ and $J_S$ to the model parameters.
For the symmetric case, $\varepsilon_f = -U/2$,
$J_S=8 V^2 /U$ and $J_A=8V^4t^2/\varepsilon^5_f$.
For the parameter sets we have compared on the two models, 
$J_A$ is smaller in the PAM than in the $d$--$p$ system. 
This implies that phase separation occurs over much larger
localization lengths in the PAM.
Indeed, 
for the conduction electron density $n_c=1/32$
we obtain $x_o \approx 28$, $21$ and $14$ for $U=6$, $5$ and $4$,
respectively. 
Because these lengths are larger than those for the $d-p$
system, we could not find definite numerical evidence of a
phase--separated state on the lattice sizes that we were able to
study; larger systems sizes would be required.
In order to confirm our picture of the origin of the phase separation,
we can instead reduce $x_0$ by increasing the hybridization $V$; we
double $V$ to $V=0.75$, 
decreasing the localization length by a factor of two to $x_o \approx 14$,
$10$ and $7$ for $U=6$, $5$ and $4$, respectively.  
Using the DMRG, we then examine $N=32$ systems with $t=0.5$,
$\epsilon_f=-U/2$  and $V=0.75$ for $U=4$, 5, and 6.
The results, presented in Fig \ref{fig07}, 
show numerical evidence for phase separation for all three $U$--values.
The chemical potential, Fig.\ \ref{fig07}(a), has a minimum as a 
function of the filling, and the conduction electron density,
Fig.\ \ref{fig07}(b), tends to localize in a region smaller than the
size of the system.
Therefore, we find that phase separation can also occur in the PAM
under the right conditions.

\begin{figure}
\begin{center}
\epsfig{width=7.5cm,file=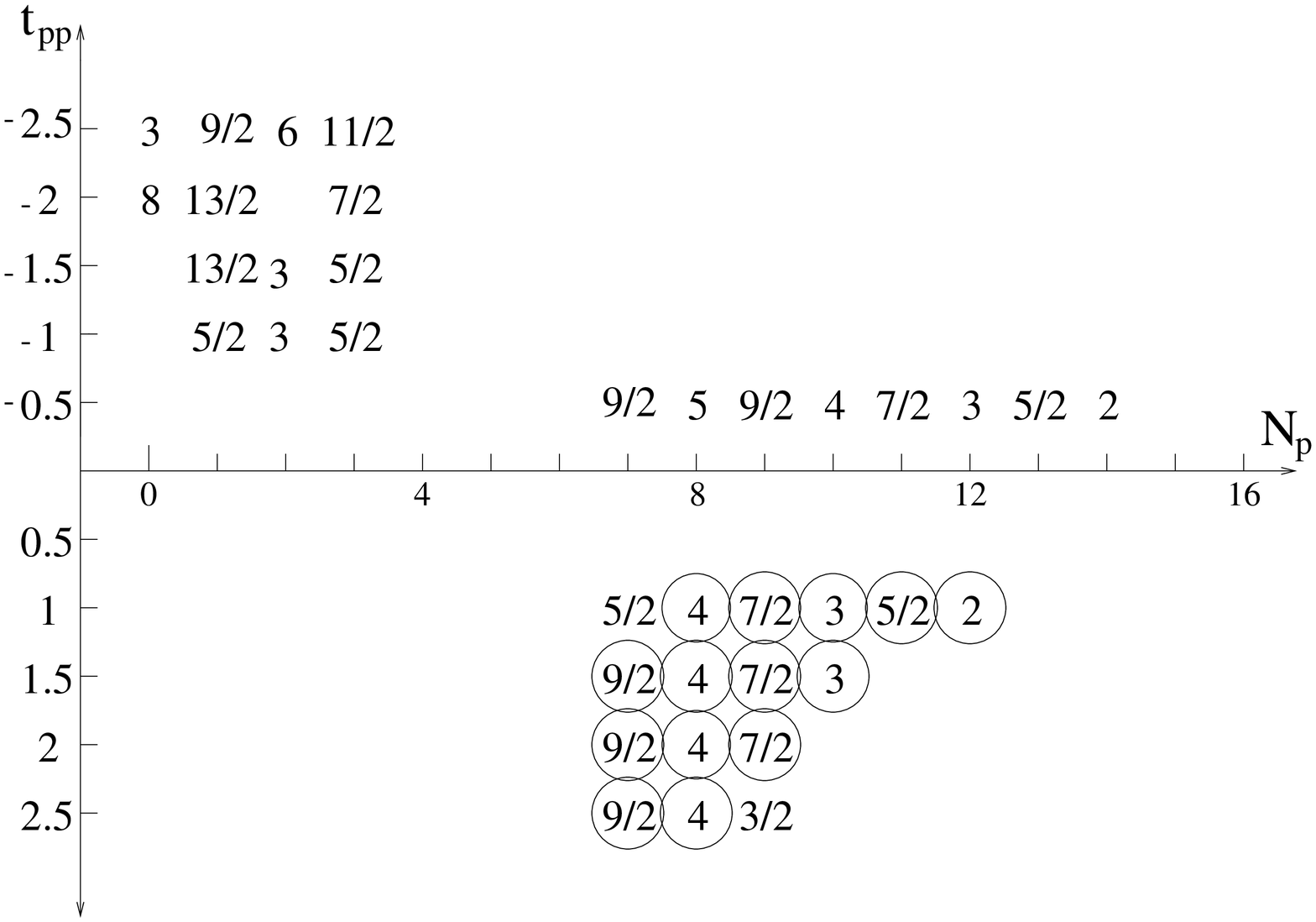}
\end{center}
\caption{Phase diagram in the plane of oxygen occupation, $N_p$, and 
hopping $t_{pp}$ for an $N=16$ copper--oxide lattice with 
$t_{pd}=1$, $U_d = 8$, $\Delta=3$, $U_p = 4$, $V_{pd} = 0.5$.
The total spin, $S$, has been calculated everywhere in the plane, but
is shown as a numerical value only when it is greater than the mimimum
value, $S=0$ for even or $S=1/2$ for odd $N_p$.
Cases of ``complete'' polarization are circled.
}
\label{fig08} 
\end{figure}

\subsection{\label{REALISTIC} Phase Diagram for ``Realistic'' Parameters}
In the second part of our work, we think of the $d$--$p$ system as the
one--dimensional analog of the copper--oxide planes in high temperature 
superconductors. 
We set the parameters $t_{pd}=1$, $U_d=8$, $U_p=4$, 
$\Delta=3$ and $V_{pd}=0.5$, values reasonable for the 
cuprates,\cite{cuprates} and then vary $t_{pp}$ and the hole filling.
The resulting phase diagram is shown in Fig.\ \ref{fig08} 
for an $N=16$ chain and both possible signs of $t_{pp}$. 
The cases with complete polarization, $S=(N-N_p)/2$, are circled. 
For $t_{pp} > 0$ the phase diagram has one ferromagnetic region
consisting almost exclusively of complete polarization for
$t_{pp}>1$ and from $n_p \sim 0.4$ to $n_p \sim 0.75$. 
The arguments of the first part of our work should also apply to this
region, i.e.\ the ferromagnetic ordering of the $f$--spins is caused
by the condensation of Zhang--Rice singlets.
As can be seen from the lack of regions of incomplete polarization, we
find no phase separation at these parameter values.
It is clear that Eq.\ (\ref{eqnnpc1}) cannot properly predict the
boundary of the ferromagnetic phase since it is invalid at the 
relatively large $N_p$ values at which we find the phase.
From Eq.\ (\ref{eqnPSlength}), the charactistic size of the
localization of the singlets is given by 
$x_0 \approx 4.5 \; t_{pp}^{1/3}$ for these parameter values. 
Our energetic arguments therefore predict no phase separation at the
densities at which we do find a ferromagnetic phase.

When $t_{pp} < 0$ the phase diagram is quite different. 
There is one point at quarter filling and $t_{pp}=-2$ for which the
ground state is fully polarized ($S=8$ for $N=16$), surrounded by a
region of smaller polarization.
Additionally, there is a region with polarization greater than
complete, but less than fully polarized for a narrow range of
$t_{pp}$ but a wide range of hole occupation, $N_p$.
More precisely, the extent of this region is 
$-0.5 \lesssim t_{pp}\lesssim -0.7$ and $7 \le N_p \le 14$.  

\begin{figure}
\begin{center}
\epsfig{width=8cm,file=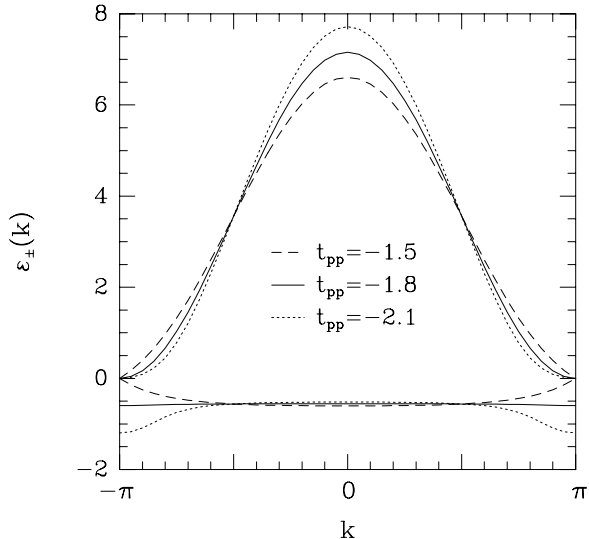}
\end{center}
\caption{
The dispersion of the non-interacting bands, $\varepsilon_{\pm}(k)$,
of the $d$--$p$ lattice for  
$t_{pd} = 1.0$, $\Delta=3.0$ and three different $t_{pp}$.
}
\label{figBands} 
\end{figure}

We first discuss the fully polarized phase at quarter filling and
$t_{pp} \approx -2.0$. 
One can gain insight into the origin of this ferromagnetic phase
by examining the noninteracting band structure.
Diagonalization of the Hamiltonian, Eq.\ (\ref{eqnCuOHam}), with 
$U_d = U_p = V_{pd} = 0$ and periodic boundary conditions leads to the
hybridized bands
\end{multicols}
\widetext
\begin{equation}
\varepsilon_{\pm}(k) = \Delta / 2 -t_{pp} \cos k \pm 
\sqrt{ (\Delta / 2 - t_{pp} \cos k )^2 + 2 t_{pd}^2 (1 + \cos k) } \; .
\label{eqnnonintbands}
\end{equation}
\begin{multicols}{2}
\narrowtext
\noindent
These bands, plotted for three $t_{pp}$ values near the fully
polarized point, are shown in Fig.\ \ref{figBands}.
As can be seen, the lower band does not overlap with the upper band
and has a dispersion that is quite flat; it becomes completely flat at
$t_{pp} = -1.78$.
Therefore, all the holes will go into the half--filled flat band at
this point.
A theorem by Mielke, \cite{mielke} subsequently extended to nearly
flat bands by Mielke and Tasaksi \cite{mielketasaki} shows rigouously
that for the Hubbard model with a low--lying completely flat band at
half--filling or less will have a fully polarized ground state.
Therefore, this region is due to flat--band ferromagnetism.
This is substantiated by the strong dependence of the polarization on
$t_{pp}$ and $n_p$; there is only a small region of partial
polarization surrounding the fully polarized point.

\begin{figure}
\begin{center}
\epsfig{width=8.0cm,file=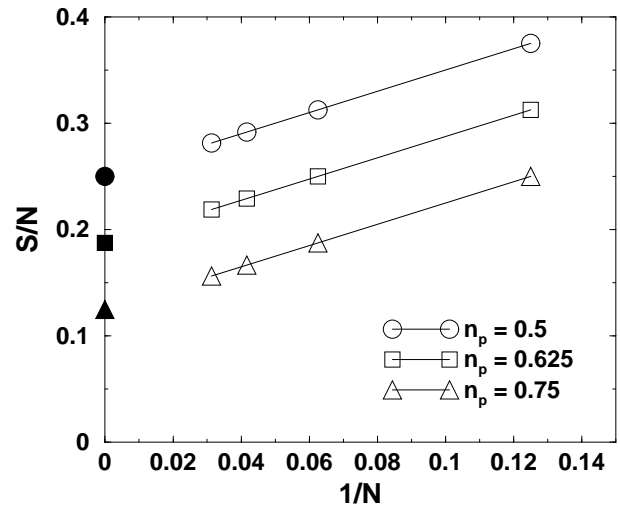}
\end{center}
\caption{Total spin per site, $S/N$, as a function of $1/N$ for a
copper--oxide lattice with
the same parameters as in Fig. \ref{fig08}, $t_{pp} = -0.5$ and three 
different fillings. 
The corresponding solid symbols show the $N\rightarrow\infty$ values
obtained from a linear extrapolation in $1/N$.
}
\label{fig11} 
\end{figure}

We next treat the narrow strip of ferromagnetism at 
$t_{pp}\approx -0.5$ and intermediate $N_p$.
The first important issue is whether the partial polarization in this
region persists in the thermodynamic limit.
In Fig.\ \ref{fig11}, 
we plot $S/N$ as a function of $1/N$ for
three different fillings.
As can be seen, $S/N$ increase linearly with $1/N$ and 
extrapolates to a finite value as $N \rightarrow \infty$.

One can gain some insight into the relationship between the sign of
$t_{pp}$ and the mechanism for ferromagnetism by examining the
behavior of small clusters.
In particular, we examine the behavior of one triangular
element of the lattice, i.e.\ a cluster consisting of one copper and
two oxygen sites, occupied by two holes ($N=1$, $N_p=1$).
This CuO$_2$ cluster can be treated analytically if spin and reflection
symmetry are taken into account:
The total spin can either be zero or one, and states
can be symmetric and antisymmetric with respect to exchange of the
oxygens.
The antisymmetric $S=0$ and the symmetric $S=0$ sectors can be
represented by $2\times 2$ matrices, the symmetric $S=0$ sector is
$4\times 4$, and the antisymmetric $S=1$ sector is $1\times 1$.
All of these matrices can be diagonalized analytically (although it is
easier, in practice, to treat the $4\times 4$ matrix numerically since
the general solution is rather complicated).
Taking $t_{pd}=1$, $U_d=8$, $U_p=4$, $\Delta=3$ and $V_{pd}=0.5$, and 
varying $t_{pp}$, we find that for $t_{pp}>0$, the ground state is always a
singlet with positive reflection symmetry.
For $t_{pp} < 0$, there is a competition between singlet formation,
favored by direct antiferromagnetic exchange, and triplet formation,
favored by third order exchange around the plaquette, which is
ferromagnetic.

\begin{figure}
\begin{center}
\epsfig{width=8cm,file=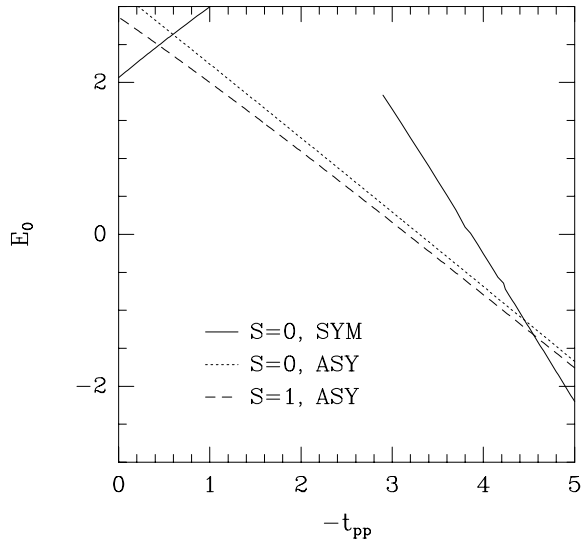}
\end{center}
\caption{ 
The energies of the three lowest lying states of the CuO$_2$ cluster
for $t_{pd}=1$, $U_d=8$, $U_p=4$, $\Delta=3$ and $V_{pd}=0.5$, as a
function of $-t_{pp}$.
Here ``SYM'' denotes a state that is symmetric with respect to exchange of
the oxygens, and ``ASY'' denotes an antisymmetric state.
}
\label{fig_E3} 
\end{figure}

In Fig.\ \ref{fig_E3}, we show the dependence of the energy of the
three lowest lying states of the CuO$_2$ cluster on $-t_{pp}$.
As can be seen, for $0 < - t_{pp} \lesssim 0.46$ and 
$-t_{pp} \gtrsim 4.5$, the ground state is a singlet with positive
reflection symmetry.
In the intermediate range, $ 0.46 \lesssim - t_{pp} \lesssim 4.5$, the
ground state is an antisymmetric triplet, which is close in energy to
the first excited state, an antisymmetric singlet.
This illustrates the effect of the sign of $t_{pp}$ on the basic
triangular element of the lattice: a localized triplet is only
possible for negative $t_{pp}$ and can occur only for a certain range
of $-t_{pp}$, when there are two holes on the plaquette.

While the triplet ground state of the CuO$_2$ cluster motivates the
possibility of a ferromagnetic state in this region, and provides an
explanation for the minimum value of $-t_{pp}$ required, it cannot
provide a complete picture of this phase.
In particular, the range of $-t_{pp}$ with a triplet ground state is
much wider in the CuO$_2$ cluster than the range of $-t_{pp}$ in the
ferromagnetic strip in Fig.\ \ref{fig08}.
When more triangles are connected together, interaction between holes
in the $p$--band can become important.
The point of maximum polarization occurs when $n_p \approx 0.5$,
i.e.\ there is one hole for every second $p$--site.
At this filling, ferromagnetism based on exchange
around the triangular elements would be somewhat frustrated since each
$p$--site is shared by two triangular elements.
These mechanisms would tend to suppress ferromagnetism.
Also, as seen in Fig.\ \ref{fig_E3}, the triplet ground state is close
in energy to a singlet excited state even on a single triangular element.

\begin{figure}
\begin{center}
\epsfig{width=8cm,file=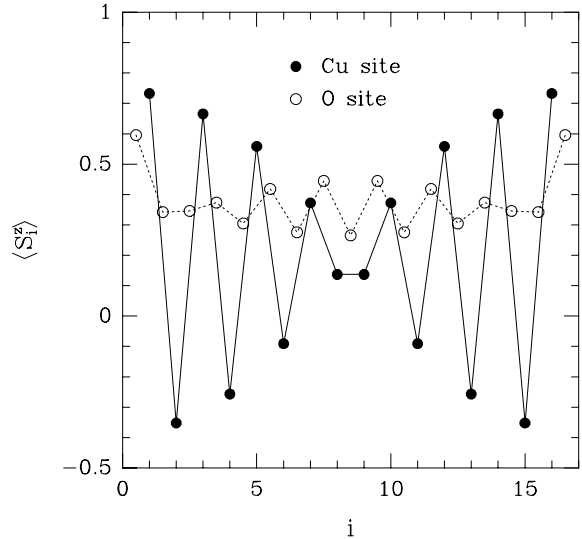}
\end{center}
\caption{
Local spin density $\langle S^z_i\rangle$, as a function of Cu--site index
$i$ for both the copper (Cu) and oxygen (O) sites on an $N=16$
$d$--$p$ lattice with $N_p = 8$ ($n_p=0.5$), $t_{pp}=-0.6$, $S_z=5$
and the remaining parameters as in Fig.\ \ref{fig08}.
For the oxygen sites, $i$ takes on half-integer values between the
corresponding Cu--sites.
}
\label{figspindenL16} 
\end{figure}

We can gain additional insight into these effects from
diagonalization on clusters larger than three sites.
For clusters of $N=2$, 3, and 4 unit cells (5, 7, and 9 total sites)
with open boundary conditions,
a ferromagnetic ground state appears at all fillings in the range 
$0 <  N_p \le N$ for $ -t_{pp} \gtrsim -t^{\rm min}_{pp}$, with 
$t^{\rm min}_{pp}$ ranging from -0.52 to -0.48.
However, the extent of this region in $t_{pp}$ and the value of the
polarization depends on $N_p$.
For $N_p = 1$, one hole in the $p$--band, and all three sizes, we obtain a
ground--state spin at or close to the maximum possible value, up to  
$t^{\rm max}_{pp}$ ranging between -2.9 and -4.2.
This behavior is like that in the $N=1$ CuO$_2$ cluster, since all the
$d$--holes are ferromagnetically polarized.
It is also reminiscent of the ferromagnet phase in the PAM,
except that it is a triplet state which delocalizes rather than a
singlet. 
Similar ferromagnetic states have been found in the one-dimensional
Kondo lattice model with ferromagnetic exchange.\cite{elbio}

For larger $N_p$, we obtain a ferromagnetic ground state up to 
$t^{\rm max}_{pp}$ ranging from -0.62 to -0.70, a value in good
agreement with that found in larger systems.
The ground--state spin varies and depends on $N_p$.
One can understand the nature of this phase by examining the
distribution of spin polarization and the spin--spin correlation
function.
We display the local spin polarization, $\langle S^z_i\rangle$, for
parameters that yield the maximum polarization on a $N=16$ cluster in
Fig.\ \ref{figspindenL16}, taking $S_z=5$, the maximal value. 
One can see that the holes on the $d$--sites are 
{\it antiferromagnetically} aligned, which we have confirmed by
examining the spin--spin correlation function.
The $p$--sites, however, all have positive polarization, and
are ferromagnetically correlated with the positively polarized
$d$--sites, i.e.\ with every other $d$--site.
The correlations between $p$--sites are generally ferromagnetic.
The picture is then that there is an alternation between triangular
plaquettes with a ferromagnetic moment and isolated $d$--holes
antiferromagnetically aligned with the neighboring plaquettes.
This phase is reminiscent of a ferrimagnet with alternating $S=1/2$
and $S=1$.
Changing the band filling can frustrate this order, changing the amount
of polarization.
One domain wall is visible in Fig.\ \ref{figspindenL16}.
Since this phase is due to a subtle competition between ferromagnetic
and antiferromagnetic exchange processes, it is not surprising that it
is quite sensitive to the value of $t_{pp}$. 
Certainly the picture of localized plaquettes breaks down as the
hopping between the $p$--sites increases.

\section{\label{CONCLU} Conclusions}

We have studied ferromagnetism in the phase diagram of the one
dimensional $d$--$p$ model. 
In the first part of this work, we have compared the phase diagram near
quarter filling of a $d$-$p$ system to the one-dimensional periodic
Anderson model. 
In both cases, we have found a ferromagnetic region near quarter filling, 
but, for equivalent values of the parameters, the
ferromagnetic region in the $d$--$p$ system is pushed towards higher
values of the Coulomb interaction and higher densities.

Between the antiferromagnetic state at quarter filling and the homogeneous
ferromagnetic region, there is a phase-separated regime in which 
ferromagnetic domains are segregated from the antiferromagnetic 
background. 
The phase--separated state arises because the loss of kinetic 
energy of the holes due to their localization in a small region is compensated
by the gain in antiferromagnetic bonds.
This regime is more evident in the $d$--$p$ system because
the antiferromagnetic coupling is stronger than in the Anderson model
and therefore the characteristic length for the ferromagnetic domains is 
smaller. 
Nevertheless, we have presented numerical evidence for the 
phase--separated state for both models. 

Recently, similar phases with a hole-rich ferromagnetic domains segregating 
from hole-undoped antiferromagnetic regions have been reported for models such
as the one-orbital ferromagnetic Kondo model and other models studied
in relation with manganites.
(See Ref. \onlinecite{elbio} and references therein.) 
In our case, it must be emphasized that the ferromagnetism is 
{\it not saturated} as in the case of these models. 
In the case of the ferromagnetic Kondo model, the phase--separated 
state was found near half-filling. 
However, it was pointed out that such a phase--separated state can arise
at low densities if an antiferromagnetic Heisenberg coupling is added
to the model.
Several studies in different numbers of dimensions using different
methods suggest that the phase--separated state is very robust. 
It is also mentioned that the inclusion of longer--range Coulomb
repulsion can turn the phase--separated state into more exotic phases,
such as striped phases with ferromagnetic domains.
While we have only studied a one-dimensional system, the arguments
presented for the physical mechanism of phase separation should still
be valid in higher dimensionsal systems.

Also, it is important to notice that the Anderson model can be
mapped to the antiferromagnetic Kondo model for large values of the
Coulomb repulsion.
It is well known that a ferromagnetic phase is also present in the
antiferromagnetic Kondo lattice model.\cite{tsunetsugu}
However, there is no phase--separated state because the
antiferromagnetic coupling is higher order than the mapping.
We believe that an additional antiferromagnetic Heisenberg
coupling of appropriate strength between the localized spins would
induce a phase--separated state in the Kondo lattice model.

In the second part of our work, we have studied the $d$--$p$ model
the parameters to values relevant to the cuprates. 
We have found that the sign of the oxygen hopping $t_{pp}$ is crucial
in determining the physics of the ground state. 
Three disconnected ferromagnetic phases were found in the phase
diagram. 
One phase occurs at positive $t_{pp}$ and is similar to the
ferromagnetic phase found in the PAM and discussed in the first part
of this work.
A second phase appears for negative $t_{pp}$ at and near quarter
filling and occurs at a point at which the occupied noninteracting
band has a flat dispersion.
This ``flat-band'' mechanism is in agreement with a picture based on
exact theorems.\cite{mielke,mielketasaki}
A third ferromagnetic phase appears for small
negative $t_{pp}$ and it is related to ferromagnetic electron
exchange processes on plaquettes with a positive product of hopping
integrals.
This phase has elements of ferrimagnetism due to an ordering of
triangular plaquettes with a ferromagnetic moment and isolated,
antiferromagnetically aligned $d$--holes.

\section*{Acknowledgments}

We thank S.\ Trugman and S.\ Daul for helpful discussions.
M. Guerrero acknowledges the support of the U. S. Department of Energy.
R.M.N acknowledges an allocation of computer time from the Centro
Svizzero di Calcolo Scientifico in Manno and support
from the Swiss National Foundation under Grant Nos. 20--46918.96 and
20--53800.98.

\end{multicols}

\end{document}